\documentclass[prd,aps,twocolumn,nofootinbib,preprintnumbers]{revtex4}
\usepackage{amsmath,amssymb,epsf}
\usepackage{graphicx}

\begin{document}

\title { \Large  \bf  On the stability of field-theoretical \\regularizations of negative tension branes}
\author{N. J. Nunes $^{1,2}$ and Marco Peloso $^2$}
\affiliation{${}^1$ William I. Fine Theoretical Physics Institute, ${}^2$ School of Physics and Astronomy \\
University of Minnesota, Minneapolis, MN 55455, USA}
\date{\today}

\begin{abstract}
Any attempt to regularize a negative tension brane through a bulk scalar requires that this field is a ghost. One can try to improve in this aspect in a number of ways. For instance, it has been suggested to employ a field whose kinetic term is not sign definite, in the hope that the background may be overall stable. We show that this is not the case; the physical perturbations (gravity included) of the system do not extend across the zeros of the kinetic term; hence, all the modes are entirely localized either where the kinetic term is positive, or where it is negative; this second type of modes are ghosts. We show that this conclusion does not depend on the specific choice for the kinetic and potential functions for the bulk scalar.
\end{abstract}

\preprint{FTPI-MINN-05/15}
\preprint{UMN-TH-2349/05}
\preprint{{\tt hep-th/0506039}}

\maketitle

\section{Introduction}

Scalar fields propagating in extra dimensions have been widely investigated. A suitable combination of bulk and brane potentials can provide a nontrivial profile for a bulk field, which can serve many purposes. For instance, it can be employed for stabilizing the size of the extra space, as in the 
Goldberger--Wise~\cite{gw} mechanism; alternatively, it can be used to localize matter in a narrow region in the extra space, as originally done by Rubakov and Shaposhnikov~\cite{rush}. Although gauge fields are not localized in Ref.~\cite{rush}, this mechanism can be considered as a field theory regularization of a delta--like lower dimensional object, as for instance a brane of string theory. The field theory description can have interesting phenomenological consequences: for instance, it allows to localize fields of different families at slightly different positions in the bulk, so to explain the fermion mass hierarchy of the standard model~\cite{ahsc}; if the scalar field can vary over cosmological times, it can allow for a greater baryon and CP violation at early times~\cite{mpst}, and overcome the obstacles for baryogenesis that characterize several models of extra dimensions~\cite{sacha}.

In these examples, the bulk scalar has standard kinetic terms. A few works discuss the possibility of more general kinetic terms. There are at least two reasons to be interested in this possibility. One is
related to the idea of self--tuning~\cite{nemanja}. In extra dimensions, a brane tension can give rise to a warping of the extra space, without inducing a $4d$ expansion~\cite{rs1}.  This typically requires a fine--tuning between the energies in the bulk and on the brane. However, one can hope that the fine--tuning can be avoided in some special cases. Ref.~\cite{nilles} presents a general theorem against this possibility, which however relies on the presence of bulk fields with standard kinetic terms. Mechanisms of self-tuning with fields with nonstandard kinetic terms were advanced for instance in Refs.~\cite{kim,holdom}. In particular, Ref.~\cite{holdom} makes use of a scalar field with a quartic derivative term.

A second motivation is the attempt to resolve a negative tension brane in field theory. The kink solution of Ref.~\cite{rush} has positive energy density, and thus it can be regarded as a resolved positive tension brane. Ref.~\cite{koka} makes use of a scalar field with the ``wrong'' sign for the kinetic term, and finds a kink profile with negative energy density. More in general, Ref.~\cite{maxim} shows that any attempt to resolve a negative tension brane through a bulk scalar necessary requires that this field is a ghost. In principle, one may hope that a ghost could also serve the purpose of self--tuning, since the negative kinetic term may absorb a brane tension to restore $4d$ flatness.~\footnote{We thank Nemanja Kaloper for drawing our attention to this possibility.}

Clearly, these proposals call for a discussion of their stability. For instance, the set-up of Ref.~\cite{koka} is clearly unstable, due to the presence of the ghost field. However, the situation is more subtle for the main model discussed in Ref.~\cite{maxim}. In this model, the kinetic term of the scalar field changes sign along the bulk. The profile of the field provides a resolution of both a positive and a negative brane, placed at two different locations in the bulk (where the kinetic term has the ``correct'' and the ``wrong'' sign, respectively). It is not obvious {\it a priori} which of these two regions controls the stability of the background. A calculation \cite{maxim}, based on an effective $4d$ potential, suggests that the model has no tachyons. However, as also acknowledged in Ref.~\cite{maxim}, the stability against ghosts requires a more accurate calculation which was beyond the aims of that work. More in general, we find very interesting to discuss the stability of models with a kinetic term which is not sign definite. This is the main purpose of the present investigation.

The stability of a given background requires a careful general relativity treatment. The fluctuations of the scalar field source and mix with the (scalar) fluctuations of the geometry. Such calculations are standard in $4d$ cosmology, for what concerns the generation of inhomogeneities and growth of structures. This framework has been extended to extra dimensional models, both for discussing the stability, and the coupling of bulk fluctuations to brane fields (with a focus on accelerator phenomenology). A general formalism can be found in Ref.~\cite{brandenberger}. The application to radion phenomenology was first done in Ref.~\cite{csaba}, and then in several other works. However, the exact identification of the physical excitations, with the bulk scalar--geometry mixing fully taken into account was performed only recently \cite{kmp}.

We generalize the computation of Ref.~\cite{kmp}, valid for a bulk field $\phi$ with standard kinetic terms, to the case of a kinetic term of the form $K \left( \phi \right) \left( \partial \phi \right)^2 \,$, where $K$ is an arbitrary function of the bulk scalar. Although we refer to Ref.~\cite{kmp} for some of the details, we try to keep the present discussion self--contained. We identify the exact $4d$ physical modes of the system.
Both the kinetic and mass term for the various modes are obtained by decomposing the original action and integrating along the compact coordinate. We explicitly show that the portion of the bulk integral where $K > 0$ gives a positive contribution to the kinetic coefficient of each mode, while the bulk region with $K < 0$ gives a negative contribution. Whether a mode is or is not a ghost then depends on where it is mostly localized.~\footnote{It is worth remarking that all these calculations are semiclassical, based on linear quantum fluctuations on a given classical background. In a path integral formulation, one has to include any type of fluctuations. In particular, one is force to consider fluctuations with arbitrarily negative kinetic energy. However, in the absence of a rigorous path integral formulation, where gravity is also included, it is worth investigating whether the model is stable or if problems arise already at the
semiclassical level.}

The linearized computation becomes questionable around the points where $K$ vanishes. At the technical level, one can expect singular terms in the equations for the perturbations. This is not obvious {\it a priori}. For instance, Ref.~\cite{finelli} discusses a somewhat similar problem in scalar field $4d$ cosmology. In that case, the equation for the metric perturbation is singular when the kinetic term for the background inflaton vanishes, $\dot{\phi} = 0 \,$. However, the equation for the Mukhanov--Sasaki~\cite{musa} variable, which properly identifies the physical mode of the system is regular. Also in the present case, one may hope that, due to the mixing with gravity, the kinetic term for the proper physical excitations may be regular where $K = 0 \,$. The computation shows that this is not the case.

We find that the singularity is mild enough that one can find normalizable modes in both the regions $K \geq 0$ and $K \leq 0\,$. However, it is also strong enough that these regions are not in communication; at the technical level, it is not possible to obtain junction conditions which relate the profile of a mode across the points where $K = 0$. We show this explicitly for the model of Ref.~\cite{maxim}; however, we also show that this a very general conclusion, irrespectively of the form of $K$. The safest interpretation is probably that the theory for the perturbations is ill defined due to the vanishing of the kinetic function, and that modifications (for instance, the introduction of higher derivative terms) are necessary to have a better defined quantum field theory. To confirm the separation of the two regions, we attempt to regularize the singularity through a cut--off, and to investigate the behavior of the solutions as the cut-off is removed. Doing so, we find that the limiting solutions either vanish where $K$ is positive or where it is negative. This regularization explicitly shows that the eigenvalue problem which determines the bulk profile of the modes effectively splits into two eigenvalue problems, characterized by a different mass spectrum. Hence, a physical mode can only live in one of the two regions; the modes which have support where $K < 0$ are ghosts, and preclude the stability of the background.

The plan of the paper is the following. In Section~\ref{one} we present the general computation for the perturbations, for a generic kinetic function $K \,$. In Section~\ref{two} we apply it to the main model of Ref.~\cite{maxim}. In Section~\ref{three} we discuss how our findings generalize beyond this application.

\section{General formalism} \label{one}

We start from the action for a scalar field $\phi$ plus gravity,
\begin{equation}
S_{\rm bulk} = \oint_0^{z_0} d^5 x \, \sqrt{- \, g} \left[ \frac{M^3}{2} \, R - \frac{K \left( \phi \right)}{2}
\left( \partial \phi \right)^2 - V \left( \phi \right) \right] \,\,,
\label{az}
\end{equation}
which is defined on a compact and periodic extra dimension $z$. For the moment, we assume
that there are no branes present, so that the calculation is simpler. We discuss below how the result changes when boundary branes are present.  We note the presence of a nonstandard kinetic term, where we allow for an arbitrary function $K \left( \phi \right)\,$. The standard case corresponds to $K = 1 \,$. More in general, $\phi$ is a ghost whenever $K < 0$. Finally, we note that a possible bulk cosmological constant is implicitly included in~(\ref{az}), by simply shifting the zero point energy of $\phi \,$.

We are interested in background solutions which only depend on $z \,$, and with the factorizable geometry
\begin{equation}
d s^2 = A \left( z \right)^2 \left[ d z^2 + \eta_{\mu \nu} d x^\mu \, d x^\nu \right] \,\,.
\end{equation}
(notice $z$ is a conformal coordinate; also, notice we have chosen the mostly positive signature for the Minkowsky metric). It is straightforward to canonically normalize the scalar field, through the relation
\begin{equation}
\varphi \equiv \int \sqrt{K} \, d \phi \,\,,
\label{psi}
\end{equation}
so that the background Einstein equations are
\begin{eqnarray}
6 \, M^3 \, \frac{A'^2}{A^2} &=& \frac{K}{2} \, \phi'^2 - A^2 \, V \,\,, \nonumber\\
\frac{A''}{A} &=& 2 \, \frac{A'^2}{A^2} - \frac{K}{3 \, M^3} \phi'^2\,\,.
\label{einbkg}
\end{eqnarray}
These two equations can be combined to give the equation of motion for $\phi$,
\begin{equation}
\phi'' + 3 \, \frac{A'}{A} \, \phi' + \frac{K'}{2 \, K} \, \phi'^2 - \frac{A^2}{K} \, V' = 0 \,\,.
\label{eomphi}
\end{equation}
Prime on $\phi$ or $A$ denotes differentiation with respect to $z \,$, while $K' \equiv d K / d \phi$, and analogously for $V$. Although~(\ref{psi}) is formally defined only where $K$ is positive, it is immediate to verify that eqs.~(\ref{einbkg})--(\ref{eomphi}) hold in general.

Periodicity conditions supplement these equations, and allow to determine the background solution. As we mentioned, we are interested in the stability of the background against scalar perturbations. We introduce the perturbations of the bulk scalar, which we denote as $\delta \phi \left( x ,\, z \right) \,$. In addition, one can reduce the system of metric perturbations to a unique mode $\Phi \left( x ,\, z \right)$, which, in the $5d$ longitudinal gauge, appears as
\begin{equation}
d s^2 = A \left( z \right)^2 \left[ \left( 1 + 2 \, \Phi \right) d z^2 + \left( 1 - \Phi \right) \eta_{\mu \nu} \,
d x^\mu \, d x^\nu \right]
\end{equation}

If $K$ is positive definite, one expects that also the calculation for the perturbations can be readily obtained from the standard one, upon the redefinition~(\ref{psi}). However, as we mentioned in the Introduction, the situation is more delicate if $K$ can change sign. As in the standard case, the two perturbations $\delta \phi$ and $\Phi$ are related by constraint Einstein equations, and only one linear combination of them is dynamical. Among different possible dynamical variables, one is particularly convenient for the diagonalization of the action; it is the $5d$ generalization of the Mukhanov--Sasaki~\cite{musa} variable $v$, introduced for the study of scalar perturbations in $4d$ cosmology. In the present case, combining the analysis of~\cite{kmp} with the redefinition~(\ref{psi}), we find
\begin{equation}
v \equiv Z \left( - \frac{\Phi}{2} - \frac{A'}{A \, \phi'} \, \delta \phi \right) \,\,,
\label{vdef}
\end{equation}
where
\begin{equation}
Z \equiv \sqrt{\vert K \vert} ~ \frac{A^{5/2} \, \phi'}{A'} \,\,.
\label{zdef}
\end{equation}

A lengthy but straightforward computation confirms that the action for the perturbations (obtained by expanding at second order in the perturbations the starting action~(\ref{az})) acquires a particularly simple form in terms of the mode $v \,$,
\begin{equation}
S^{\left(2\right)} = \frac{1}{2} \oint d^5 x \, \sigma \! \left( K \right) \, v \left[ \Box + \frac{d^2}{d z^2} - \frac{Z''}{Z} \right] v \,\,,
\label{az2}
\end{equation}
where $\sigma \!\left( K \right)$ denotes the sign of $K \,$.  It is worth noting that the use of the generalized Mukhanov--Sasaki variable automatically ``rescales away'' the function $K$ from the kinetic term, up to its sign. However, the potential problems with vanishing $K$ are ``encoded'' in the fact that $Z$ vanishes for $K = 0$, so that, unless there are cancellations with $Z''$, the effective potential for the perturbations is divergent. We already discussed this problem in the Introduction; in the next Section we will discuss how this problem can be dealt with in a specific example. Here, we simply proceed with the computation, by decomposing the $5d$ variable $v$ into KK modes,
\begin{equation}
v \left( x , z \right) = \sum_n {\tilde v}_n \left( z \right) \, Q_n \left( x \right) \,\,.
\label{deco}
\end{equation}
The modes $Q_n$ represent quantum fields in the $4d$ description of the model, while ${\tilde v}_n$ are the corresponding wave functions in the bulk. They are determined by separating the equation of motion for $v$ which follows from~(\ref{az2}),
\begin{equation}
\left(\frac{d^2}{dz^2} -\frac{Z''}{Z} + m_n^2\right){\tilde v}_n = 0 \,\,.
\label{eigenv}
\end{equation}
This eigenvalue equation, together with the periodicity condition along the extra dimension, determines the bulk profiles ${\tilde v}_n \,$, as well as the spectrum of the theory, $\left\{ m_n^2 \right\} \,$. The exact profile found here is then employed to compute whether a mode is a ghost or not, as we show now.

As remarked in the Introduction, we are interested in the kinetic terms of the $4d$ modes. Due to the hermiticity of~(\ref{az2}), different modes are orthogonal, and~(\ref{az2}) separates into the sum of decoupled quadratic actions for each mode,
\begin{equation}
S^{\left(2\right)} = \sum_n S^{\left(2\right)}_n = \frac{1}{2} \, \sum_n C_n \int d^4 x \, Q_n \left[ \Box - m_n^2 \right] Q_n \,\,,
\end{equation}
where (comparing with~(\ref{az2})) it is immediate to see that the coefficients $C_n$ are given by
\begin{equation}
C_n = \oint d z \, \sigma \left( K \right) \, {\tilde v}_n^2 \,\,.
\label{norma}
\end{equation}
A further rescaling ${\hat Q}_n \equiv Q_n / \sqrt{ \vert C_n \vert}$ renders the mode canonically normalized. However, the sign of $C_n$ is not rescaled away by this redefinition, and it thus controls whether the mode ${\hat Q}_n$ is a ghost or not. As we see, each coefficient $C_n$ gets positive contributions where $K > 0$, and negative contributions where $K <0$. So, the nature of a mode is determined by whether its wave function is mostly localized where $K > 0$, or $K<0$.

For completeness, we conclude this section by giving the result of the computation when also boundary terms are present. More specifically, the extra coordinate is assumed to lie on the compact interval $0 < z < z_0$, delimited by two orbifold  ($Z_2$ symmetric) boundary branes. We assume that the brane contains a $\phi-$dependent potential term,
\begin{equation}
S_{{\rm brane},i} = -\int d^4 x \sqrt{ - \gamma_i} \left\{ 2 \, M^3 \, \left[ K \right] + U \left( \phi \right) \right\}_i \,\,,
\end{equation}
where $\gamma_i$ is the induced metric at the brane location, while $\left[ K \right]$ denotes the jump of the trace of the extrinsic curvature across the brane. In principle, kinetic terms for $\phi$ could be also present at the boundaries, and they would have the effect of modifying the kinetic term for the $4d$ modes. However, we will not consider them here.

The branes enforce boundary conditions, which replace the periodicity conditions considered so far. For the background, we have
\begin{equation}
U_i = \mp \, 6 \, M^3 \frac{A'}{A^2} \Big\vert_i \;\;\;,\;\;\; U_i' = \pm \frac{2 \, K \, \phi'}{A} \Big\vert_i \,\,,
\end{equation}
where the upper/lower sign refer to the brane at $z = 0 / z_0\,$, respectively. For the perturbations we get instead
\begin{eqnarray}
&& \Phi' + 2 \, \frac{A'}{A} \, \Phi - \frac{2 \, K \, \phi'}{3 \, M^3} \, \delta \phi = 0 \,\, \nonumber\\
&& \delta \phi' - \phi' \, \Phi \mp \frac{A}{2 \, K} \, U'' \, \delta \phi + \frac{K' \, \phi'}{K} \delta \phi = 0 \,\,.
\label{jpert}
\end{eqnarray}
These new equations, together with the eigenvalue equation~(\ref{eigenv}), allow to determine the spectrum and the bulk profiles of the modes. A simple extension of the calculation of~\cite{kmp} shows that the coefficients $C_n$ acquires a boundary contribution
\begin{equation}
C_n = \int d z \, \sigma \left( K \right) \, {\tilde v}_n^2 - \frac{3 \, M^3 \, A^4}{4 \, A'} \, {\tilde \Phi}_n^2 \Big\vert_0^{z_0} \,\,,
\end{equation}
where we have defined $\Phi = \sum_n {\tilde \Phi}_n \, Q_n \,$, in strict analogy to~(\ref{deco}). The boundary values of ${\tilde \Phi}_n$ can be obtained from the ones of ${\tilde v}_n$ through the definition~(\ref{vdef}) and the boundary conditions~(\ref{jpert}).

\section{Application} \label{two}
Let us discuss how the above formalism applies to the model of Ref.~\cite{maxim}.
The model is characterized by the bulk action~(\ref{az}), with
\begin{eqnarray}
K \! \left( \phi \right) &=& 3 \, M^3 \, A_0 \, \frac{\phi / \phi_0}{\phi_0^2 - \phi^2} \,\,,
\nonumber \\
V \! \left( \phi \right) &=& - 6 \, M^3 \, w^2 \, A_0^2 \left\{ 1 - \frac{1}{4 \, A_0} \, \frac{\phi}{\phi_0} - \frac{\phi^2}{\phi_0^2} \right\} \,\,,
\label{maxmodel}
\end{eqnarray}
extending on a periodic interval ($M^3$ is the five dimensional Planck mass while
$A_0 ,\, w ,\, \phi_0$ are some constants). The background solution is particularly simple and elegant. In normal coordinates, defined as
\begin{equation}
d s^2 = d y^2 + A \left( y \right) \eta^{\mu \nu} d x_\mu d x_\nu \,\,,
\end{equation}
one has
\begin{eqnarray}
\phi \left( y \right) &=& \phi_0 \, \cos \left( w \, y \right) \,\,, \nonumber\\
A \left( y \right) &=& \exp \left[ A_0 \, \cos \left( w \, y \right) \right] \,\,.
\label{back}
\end{eqnarray}

\begin{figure}[h]
\centerline{
\includegraphics[width=0.5\textwidth]{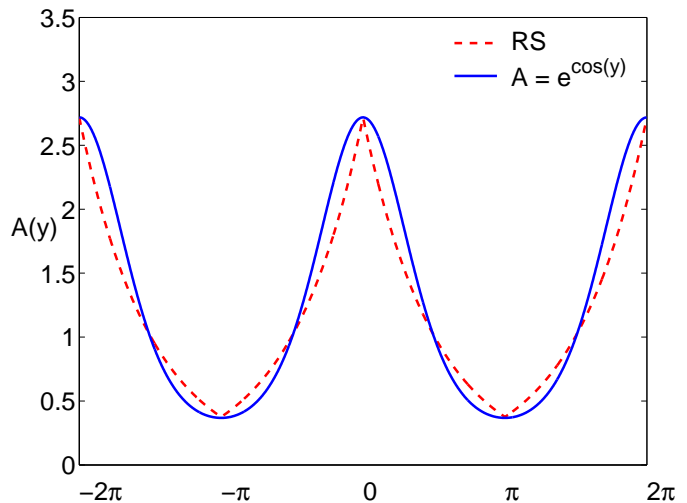}
}
\caption{Dashed line: typical $A \sim \exp \left( - \vert y \vert \right)$ bulk profile for a bulk cosmological constant. Branes are necessary at the extrema of $A$, to match the different patches. Solid line:
regularized $A \sim \exp \left(  \cos \, y\right)$ profile for the model~(\ref{maxmodel}).
}
\label{fig:fig1}
\end{figure}

The geometry is characterized by a nontrivial periodic evolution of the warp factor $A \,$. The sourcing  bulk scalar is also periodic in the bulk, and the scalar density $\phi^2$ is mostly localized at $y = 0 \,, \pi / w \,$, where $A$ has its extrema. This model represents an attempt to regularize a brane/anti--brane system. This is shown in fig.~\ref{fig:fig1}, where we compare the background~(\ref{back}) with the Randall--Sundrum background~\cite{rs1}. In the present case, the maximum (minimum) of $A$ is due to a delocalized field rather than to a positive (negative) tension brane. As shown in~\cite{maxim}, a bulk scalar field with the standard sign for the kinetic term cannot give rise to a minimum of $A$, irrespectively of its potential. Indeed, we observe from the definitions of $K(\phi)$ and $\phi(y)$ that $K$ is positive in the interval $0\le w\,y < \pi/2$ close to the maximum of $A(y)$, whereas it is negative in the region $\pi/2 < w\,y \le \pi$ near the minimum of the warp factor where $\phi$ regularizes the negative tension brane.

Let us now compute the scalar perturbations around the background~(\ref{back}). It is convenient to rewrite eq.~(\ref{eigenv}) in normal coordinates. In terms of the rescaled variable $\psi \equiv A^{1/2} \, {\tilde v} \,$ (for shortening the notation, we omit the subscript $n$) one finds
\begin{equation}
\frac{d^2 \, \psi}{d y^2} + {\rm e}^{-2 \, A_0 \, \cos \left( w \, y \right)} m^2 \, \psi - V_{\rm eff} \, \psi = 0 \,\,,
\label{sch}
\end{equation}
where
\begin{eqnarray}
&& V_{\rm eff} \equiv \frac{w^2}{4} \Bigg[ \cot^2 \left( w \, y \right) + \frac{8}{\sin^2 \left( w \, y \right)} - \frac{4}{\sin^2 \left( 2 \, w \, y \right)} + \nonumber\\
&& \quad\quad\quad\quad\quad\quad + \frac{8 \, A_0}{\cos \left( w \, y \right)} + 16 \, A_0^2 \, \sin^2 \left( w \, y \right) \Bigg] \,\,.
\label{exact}
\end{eqnarray}
\begin{figure}[h]
\centerline{
\includegraphics[width=0.5\textwidth]{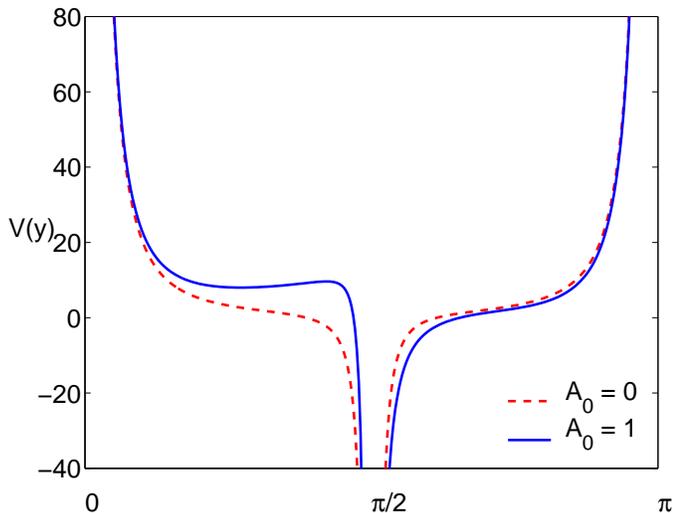}
}
\caption{Effective potential for the $4d$ modes, eq.~(\ref{exact}), as a function of the bulk variable $w \, y\,$. The locations $w \, y = 0, \pi \,$, correspond to the regularized positive and negative tension brane, respectively. The kinetic function $K$ vanishes at $w \, y = \pi /2 \,$.}
\label{fig:fig2}
\end{figure}

It is always possible to rescale $y$ such that $w = 1 \,$. We do so from now on. The effective potential is $Z_2$ symmetric around $y=0$ (``regularized'' positive brane) and $y=\pi$ (``regularized'' negative brane), where it diverges to $+ \infty \,$ (see fig.~\ref{fig:fig2}). In addition, it is unbounded from below where the kinetic function~$K$ vanishes. Despite of the singularities, we will now show that~(\ref{sch}) admits solutions which are finite everywhere. However, there is an intrinsic ambiguity in matching the solutions across the singularities of $V_{\rm eff}$. To see this, let us first solve~(\ref{sch}) for $A_0 = 0 \,$, and then for small $A_0 \,$ (although the model~(\ref{maxmodel}) is trivial for $A_0 = 0$, the formal problem~(\ref{sch}) is still defined).

\vspace{1cm}
For $A_0 = 0$, the potential is symmetric also around $y = \pi /2$; in the interval $0 \leq
y \leq \pi / 2 \,$, we approximate it as
\begin{eqnarray}
V_{\rm eff} \simeq \left\{ \begin{array}{c}
\frac{2}{y^2} + \frac{1}{6} \quad,\quad\quad 0 \leq y \leq 1 \,\,, \\ \\
- \frac{1}{4 \left( y - \frac{\pi}{2} \right)^2} + \frac{5}{3} \quad,\quad\quad 1 \leq y \leq \frac{\pi}{2} \,\,.
\end{array} \right.
\label{approx}
\end{eqnarray}
This approximated form reproduces the potential where it diverges (up to terms which vanish as $y \rightarrow 0,\, \pi/2$), and it is very close to the exact potential everywhere. We verified that the numerical solutions to the exact problem~(\ref{exact}) are very well approximated by the analytic ones of~(\ref{approx}). The ``matching point'' $y =1$ has been chosen by comparison with the exact form of the potential, and we verified that the solutions do not change significantly if the matching point is slightly moved away from $1$. 

Equation (\ref{sch}), with the approximated potential~(\ref{approx}), is solved by
\begin{eqnarray}
\label{modes}
\psi &=& C_1 \left[ \cos \left( \alpha \, y \right) - \frac{\sin \left( \alpha \, y \right)}{\alpha \,  y} \right] \\
&&+ \, C_2 \left[ \sin \left( \alpha \, y \right) + \frac{\cos \left( \alpha \, y \right)}{\alpha \, y} \right]  \;\;,  \quad\quad\quad 0 \leq y \leq 1 \,\,, \nonumber\\
\psi &=& C_3 \, \sqrt{\xi} \, J_0 \, \left( \beta \, \xi \right) + C_4 \, \sqrt{\xi} \, Y_0 \left( \beta \, \xi \right) \;\;,
\quad 1 \leq y \leq \frac{\pi}{2} \,\,, \nonumber
\end{eqnarray}
where in the last line we have defined $\xi \equiv \pi / 2 - y \,$, and where
\begin{equation}
\alpha \equiv \sqrt{m^2 -\frac{1}{6}} \quad\quad,\quad\quad \beta \equiv \sqrt{m^2 - \frac{5}{3}} \,\,.
\end{equation}
Since the potential is symmetric around $\pi/2$, in the range $\pi/2 \leq y \leq \pi$ the solution 
can be written as~(\ref{modes}), upon the substitution $y \rightarrow \pi - y \;,\;\; C_i \rightarrow D_i \,$.

Let us now identify the unknown quantities, and see whether we can determine them by boundary conditions (continuity of $\psi$ and $\psi'$ at $y=1,\, \pi/2 ,\, \pi - 1 \,$). In the first interval, the mode proportional to $C_2$ diverges at the origin. This immediately sets $C_2 = 0 \,$. Analogously, regularity 
at $\pi$ sets $D_2 = 0 \,$. We are thus left with $7$ unknowns: the six coefficients $C_{1,3,4}$, $D_{1,3,4} \,$, and the eigenvalue $m^2 \,$. One coefficient cannot be determined, since the overall normalization of the mode cannot be obtained from the linearized calculation we are performing (the normalization is fixed as explained in the previous section). Hence, we can fix here $C_1 = 1 \, $.

We are thus left with $6$ unknowns, and with the request of continuity of $\psi$ and its derivative at the three locations $y = 1 ,\, \pi/2 ,\, \pi - 1 \,$. The modes are regular at $y = 1 ,\, \pi - 1\,$, so we have four nontrivial boundary conditions there. As can be expected, the problems arise for the matching at $y = \pi / 2 \,$. At this point, the mode~(\ref{modes}) vanishes, while its derivative diverges,
\begin{equation}
\psi \sim C_3 \, \sqrt{\xi} + \frac{2 \, C_4}{\pi} \, \sqrt{\xi} \, \ln \, \xi \quad,\quad\quad {\rm as} \;\; \xi = \frac{\pi}{2} - y \rightarrow 0 \,\,.
\end{equation}
(and analogously for $y$ approaching $\pi /2$ from the right). Hence, the two boundary conditions at $\pi / 2$ are absent, and a global solution cannot be determined. This confirms the expectation that the linearized problem is ill posed, due to the vanishing of the kinetic function $K \,$.

One can expect that higher order terms can be relevant where $K$ vanishes, providing a cut--off where the effective potential is unbounded. In the following we regularize the potential by hand with a cut--off at $y \sim \pi / 2 \,$ (which allows to solve the linearized problem).  More precisely, we take the potential to be constant (matching the value from~(\ref{approx})), in the interval $\pi / 2 - \epsilon \leq y \leq \pi / 2 + \epsilon \,$; we then study the behavior of the solution as $\epsilon \rightarrow 0 \,$. Fortunately, this procedure leads to a well defined and normalizable solution. This limiting solution is the one which could have been easily guessed from elementary quantum mechanics. The coefficients of the modes which are more divergent as $y \rightarrow \pi / 2$, namely $C_4$ and $D_4$, vanish. The first eigenvalues are approximately given by (in units of $w^{-2} \,$)
\begin{equation}
m^2 \sim 6 ,\, 20 ,\, 42 ,\, 72 ,\, \dots
\label{eigen}
\end{equation}
As always for symmetric potentials, each eigenvalue $m^2$ admits two degenerate solutions, one symmetric and one antisymmetric around $\pi / 2 \,$. We confirmed these solutions through a fully numerical calculation~\footnote{The numerical calculation is a boundary value problem, and it can be solved with a shooting method. See~\cite{kmp} for details.} using the exact potential~(\ref{exact}). Although modes with negative $m^2$ are in principle a possibility, the numerical analysis does not reveal indications for their existence.

\vspace{1cm}
The more relevant case of $A_0 \neq 0$ can be studied analogously. The additional terms in the potential can be also approximated by simple polynomials, and analytic solutions can be obtained in terms of new special functions. Alternatively, the problem can be studied numerically for any given value of the cut--off, and one can verify that also in this case limiting solutions are reached when $\epsilon$ is sent to zero. The results are quite interesting, since -- for the reasons we will now argue -- the limit $A_0 \rightarrow 0$ is not continuous.

\begin{figure}[h]
\centerline{
\includegraphics[width=0.5\textwidth]{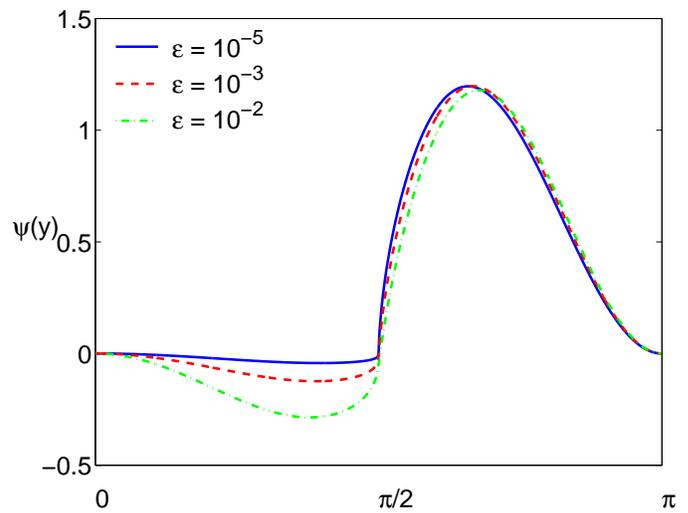}
}
\caption{A particular mode, for $A_0 = 0.1 \,$, and for three different values of the cut--off $\epsilon \,$. The normalization is here fixed by requiring $\int \psi^2 d y = 1$ for all the cases. The limiting solution (for $\epsilon \rightarrow 0$) has support only in the interval $\pi / 2  \leq y \leq \pi \,$, where the mode is a ghost.
}
\label{fig:fig3}
\end{figure}

For small $A_0$ the eigenvalues occur in finely split pairs. This is due to the fact that the $A_0$ part of the potential~(\ref{exact}) is not symmetric around $\pi / 2 \,$. As $A_0 \rightarrow 0 \,$, the values~(\ref{eigen}) are recovered; however, the limiting solutions are not any longer symmetric or antisymmetric around $\pi / 2 \,$. For any non-vanishing $A_0$ (not necessarily small), the limiting solutions split in two groups: one characterized by modes which are nonvanishing only in the interval $0 \leq y \leq \pi/2 \,$ (group I), and one by modes which are nonvanishing in the complementary interval $\pi/2 \leq y \leq \pi \,$ (group II). This can be easily explained. As we have seen, for $\epsilon = 0 $ the two halves of the space are not in communication. Hence, the eigenvalue problem~(\ref{sch}) effectively decouples in two different problems. Unless the potential is symmetric in the two halves, the two eigenvalue problems admit different eigenvalues. It is then impossible for an eigenmode to have support in both halves. The decoupling is visible in the $\epsilon \rightarrow 0$ limit, as shown for a particular case in fig.~\ref{fig:fig3}.

Let us finally return to the initial question, whether a $4d$ mode is a ghost or not. Solutions in group I have support in the interval $0 \leq y \leq \pi / 2 \,$, where $K$ is positive. The normalization coefficient $C_n$ for these modes, see eq.~(\ref{norma}), is positive, and hence they are well behaving $4d$ scalars. On the contrary, the modes which have support in the other half of the bulk, where $K$ is negative, are ghosts. Although we have investigated only a limited set of values for $A_0$, this second class of models has always been present for all the values we have considered (in a comparable amounts to the modes in group I). We therefore conclude that the model~(\ref{maxmodel}) is unstable already at the semiclassical level.

\section{Discussion} \label{three}

Our main motivation was to study explicitly how the vanishing of the kinetic function $K$ affects the
system of perturbations. Quite in general, one can expect singularities in correspondence to the zeros of $K \,$. We found that, for the specific model considered in Ref.~\cite{maxim}, the singularities are mild enough so that the regions where $K \leq 0$ and $K \geq 0$ admit normalizable modes for the perturbations; however, we also saw that the singularity is strong enough so that these regions do
not communicate, and the modes have support only in one of them. Modes which have support where $K$ is negative are ghosts, and preclude the stability of the background. We now argue that this effect is quite general, irrespectively of the detailed form of $K \,$. To see this, assume that $K$ vanishes at some given point $y_* \,$ in the bulk. In general, we can expand,
\begin{equation}
K \sim \left( y - y_* \right)^\alpha \equiv \xi^\alpha \quad,\quad\quad \alpha > 0
\end{equation}
at small $\xi \,$. Using the background eqs.~(\ref{einbkg}) and (\ref{eomphi}) to determine the functional form of $\phi'$ and $A'$ where $K$ vanishes, we find that, in a neighborhood of $y_* \,$,
\begin{equation}
Z \sim \Bigg\vert \frac{K \, V'}{V \, K'} \Bigg\vert^{1/2} \sim \Bigg\vert \frac{K \: \partial V / \partial \xi}{V \: \partial K / \partial \xi} \Bigg\vert^{1/2} \,\,.
\end{equation}
If the potential $V$ is regular and nonvanishing at $y_* \,$, we then find
\begin{equation}
Z \sim \xi^{1/2} \quad\quad \Rightarrow \quad\quad V_{\rm eff} \sim \frac{Z''}{Z} \sim \xi^{-2} \,\,.
\label{general}
\end{equation}
Recalling the expansion~(\ref{approx}), we see that the model~(\ref{maxmodel}) is not exception to the general rule, and that the degree of divergency of the effective potential is typically $-2 \,$. Hence, the problems and the instability found for this model are expected to be a general issue, whenever the kinetic function is allowed to vanish.

As manifest in~(\ref{general}), a possible ``cure'' to the instability could be to arrange $V$ to vanish precisely where $K$ also does. This is a rather trivial solution, which however is likely to postpone the problem at orders higher than quadratic. More in general, one may hope that higher orders in $\partial \phi$ may allow the theory to have a better behavior where the quadratic term vanishes.

\vspace{0.9cm}
\centerline{\bf Acknowledgements}
\vspace{0.1cm}

We thank Nemanja Kaloper, Stefan Groot Nibbelink and Maxim Pospelov for very useful discussions. This work has been supported in part by the Department of Energy under contract DE-FG02-94ER40823
at the University of Minnesota.

\end{document}